\documentclass[preprint,prc,aps]{revtex4}

\usepackage{amsmath} 
\usepackage{amssymb} 
\usepackage{epsfig}
\usepackage{pstricks}
\usepackage{ulem}

\begin{document}
\title{An inspection on the Borel masses relation used in\\ QCD sum rules}
\author{M. E. Bracco}
\affiliation{Faculdade de Tecnologia, Universidade do Estado do Rio de Janeiro, 
Rod. Presidente Dutra Km 298, P\'olo Industrial, 27537-000 , Resende, RJ, Brazil.}

\author{M. Chiapparini and B. Os\'orio Rodrigues}
\affiliation{Instituto de F\'{\i}sica, Universidade do Estado do Rio de 
Janeiro, Rua S\~ao Francisco Xavier 524, 20550-900 Rio de Janeiro, RJ, Brazil. }
\begin{abstract}
 In this work, we studied the Borel masses relation used in QCDSR calculations. These masses are the parameters of the Borel transform used when the three point function is calculated. We analised an usual and a more general linear 
relations. We concluded that a general linear relation between these masses provides the best results regarding the standard deviation.
\end{abstract}

\maketitle

\section{Introduction}
Hadrons are thinking as bound states of quarks. This problem it is not simple due that asymptotic 
freedom and confinement shows that the underlying theory, QCD, must have a complex structure.
The breakdown of asymptotic freedom is manifest by the power corrections that are introduced due to the non-perturbative effects of the QCD vacuum. The QCD sum rules (QCDSR) method provides a way to extract information from the ultraviolet regime (asymptotic freedom) and give us information about bound states and hadronic quantities such masses, coupling constants, form factors and so on \cite{1979NuPhB.147..385S}. In order to  pick out the lowest lying resonance in a particular channel, we define moments by taking derivatives of the correlation function in the dispersion representation. In principle, a large mass scale $Q^2$ makes all the corrections  small and the derivatives can be taken. In this way, a new variable $M^2$ is introduced instead of the moment. This procedure corresponds to introduce the Borel transform in the calculation. To obtain the form factors and coupling constants for a particular process, it is used the three point correlation function, which requires a double Borel transformation in order to apply the QCDSR technique. This double Borel transform introduces two Borel masses as two parameters, instead of the original momenta.

In order to exemplify, we consider the $D^* D \rho$ vertex, with $\rho$ off-shell, and the $B_s^*BK$ vertex, with $B$ off-shell. After performing the two Borel transformations we get the sum rule to obtain the form factor. In the case of the $D^* D \rho$ vertex we obtain the following expression for the form factor:  
\begin{align}
g^{(\rho)}_{D^* D \rho}(Q^2) = \frac{- \frac{1}{4\pi^2} \int^{s_{sup}}_{s_{inf}} ds\int^{u_{sup}}_{u_{inf}} du\:DD[\Pi^{QCD}] e^{-s/M^2} e^{-u/M'^2}}{\frac{f_{\rho} f_{D^{*}} f_{D} m_{\rho} m_{D^{*}} \frac{m^2_{D}}{m_c}}{(m_{\rho}^2 + Q^2)} e^{-m^2_{D^*}/M^2}e^{-m^2_D/M'^2}}\;,
\label{eq:fatordeformarho}
\end{align}
 while for the $B_s^*BK$ vertex the form factor is:
\begin{align}
 g^{(B)}_{B^*_sBK}(Q^2)=\frac{-\frac{1}{4 {\pi}^2}\int ds\int du\: DD[\Pi^{QCD}]
e^{-s/M^2}e^{-u/{M^{\prime}}^2}}{\frac{2f_B  f_K f_{B^*_s} m_B^2
m_{B^*_s}^2}{
(m_b+m_s) (Q^2+m_B^2)}e^{-m^2_K/{M^{\prime}}^2} e^{-m^2_{B^*_s}/M^2}}\;.
\label{eq:fatordeformaBBK}
\end{align}
The determination of the form factor depends on two Borel masses $M^2$ and $M'^2$ and two continuum thresholds, $s_{sup}$ and $u_{sup}$. These thresholds are given by $s_{sup} = (m_i + \Delta_s)^2$ and $u_{sup} = (m_o + \Delta_u)^2$, where $m_i$ and $m_o$ are the masses of the incoming and outcoming on-shell mesons respectively, and the quantities $\Delta$ are adjustable parameters, of the order of the energy gap  between the ground and the first excited states. 

\begin{figure}\centering
\resizebox{0.95\textwidth}{!}{\includegraphics{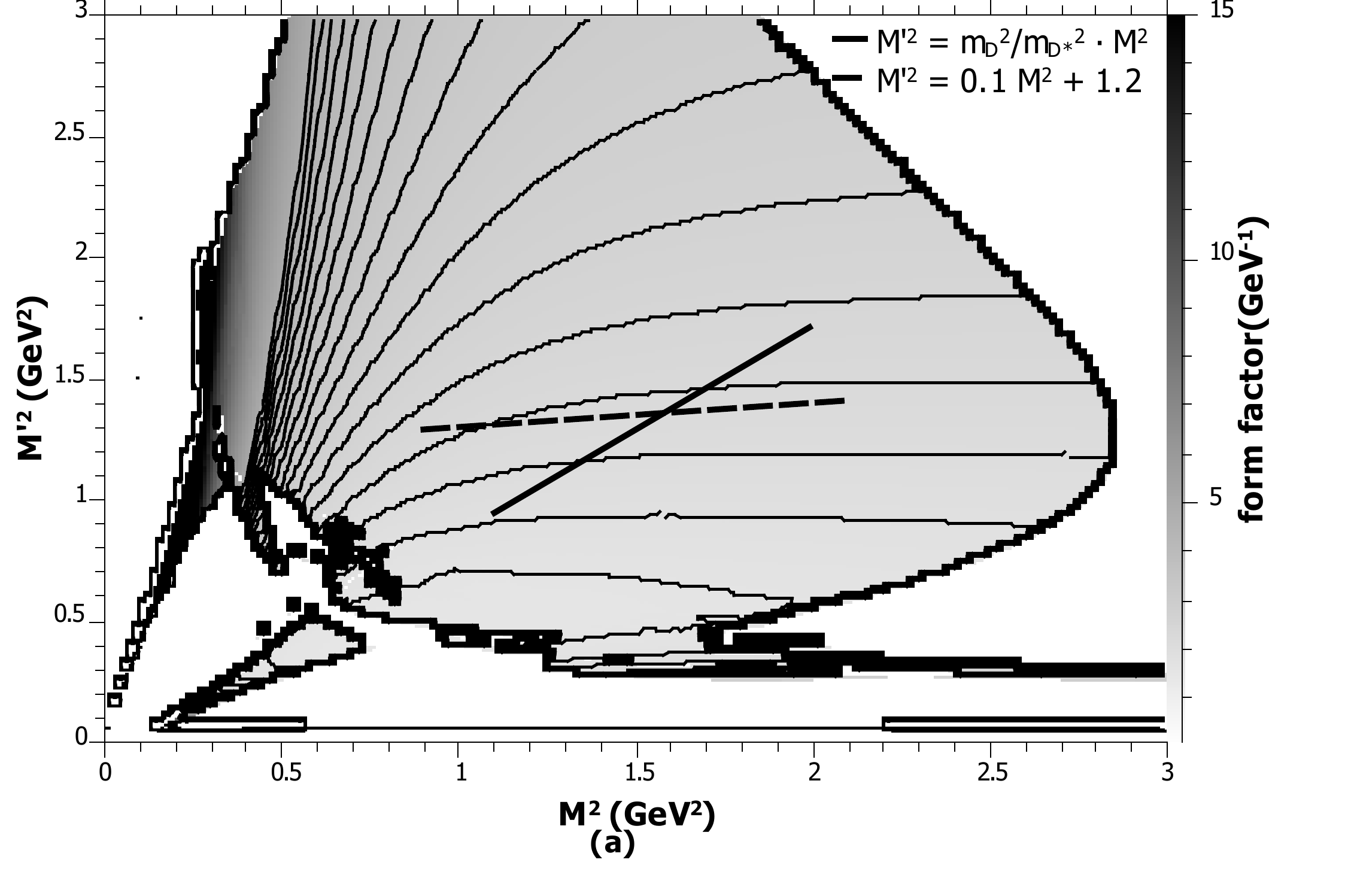}\includegraphics{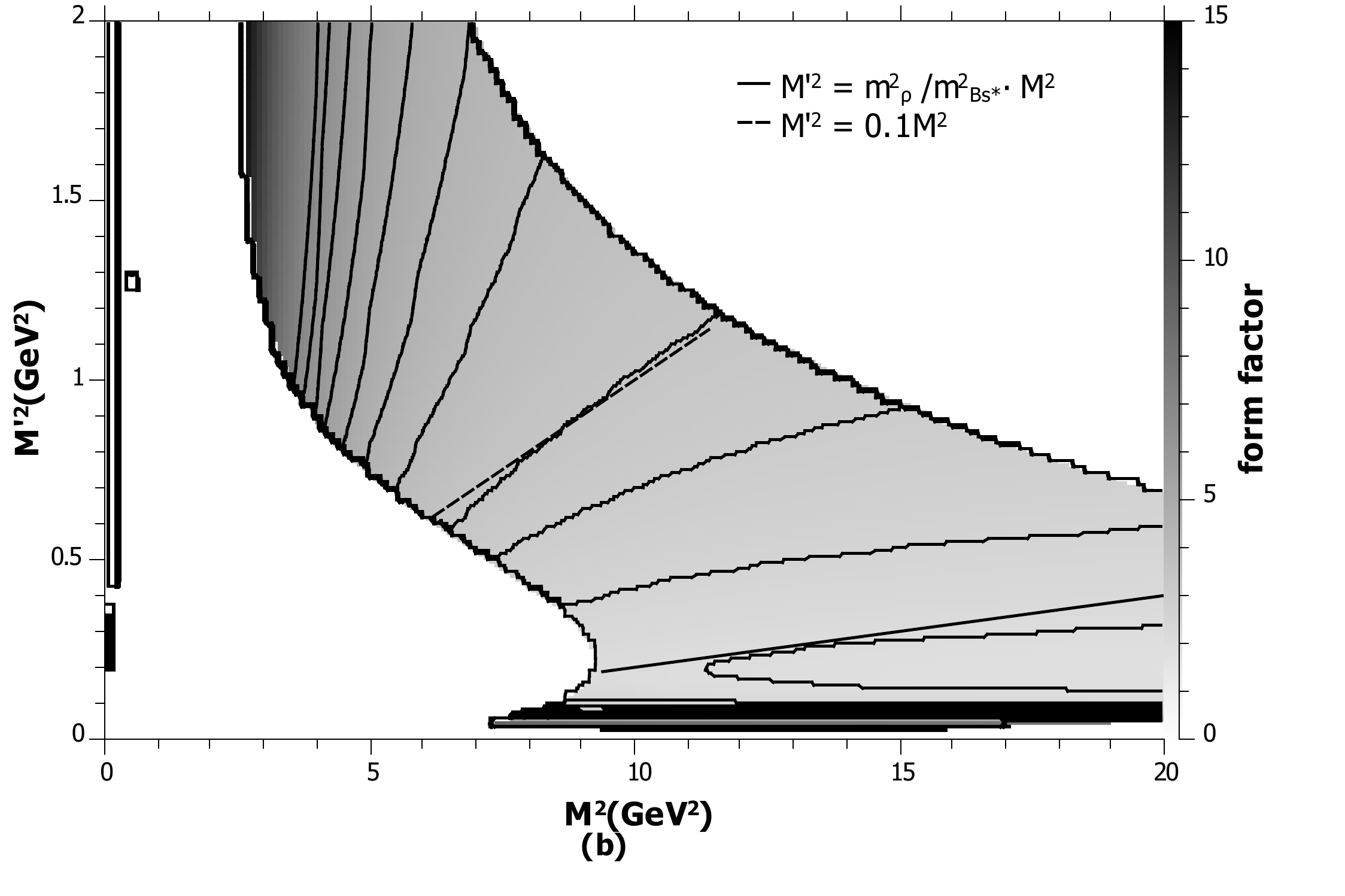}}
\caption{Borel masses region for the $D^* D \rho$ vertex (panel \mbox{(a)}, 
with $\Delta_s = 0.3$ GeV and $\Delta_u = 0.6$ GeV) and for the  $B_s^* B K$ vertex (panel 
\mbox{(b)}, with $\Delta_s = 0.5$ GeV 
and $\Delta_u = 0.6$ GeV), for $Q^2 = 2$ GeV$^2$.}\label{fig1}\end{figure}

 In order to perform an analysis of the optimal choice for the Borel masses and continuum thresholds, we solved numerically Eqs.(\ref{eq:fatordeformarho}) and (\ref{eq:fatordeformaBBK}) for different values of $Q^2$. We obtained the thresholds in an auto-consistent way and we plotted the Borel mass squared $M^2$ versus $M'^2$. The pairs $(M^2, M'^ 2)$, uncorrelated in principle, should be chosen in order to obtain a good sum rule. Two conditions define a good sum rule:
\begin{itemize}
	\item[] (i) The pole contribution to the QCD correlation function be always greater than the continuum contribution. This give the upper limit of the Borel mass.
	\item[] (ii) The contributions to the QCD correlation function from the high order condensates must be smaller than the leading (perturbative) contribution. Here a lower limit of the Borel mass is chosen. This  guarantees the convergence of the QCD correlation function series, allowing its truncation.
\end{itemize}
The pairs $(M^2,M'^2)$ satisfying the above conditions define a finite region in the real plane. This region is showed in Fig.\ref{fig1} for the $\rho$ off-shell case of the $D^* D \rho$ vertex (panel (a)) and for the $B$ off-shell case of the $B_s^* B K$ vertex (panel (b)).
At this point we could ask the following: what are the optimal pairs $(M^2,M'^2)$ which allow us to have a stable form factor? 

Given a set of values of $(M^2,M'^2)$, we calculate the mean value of the form factor and its standard deviation inside the set. Small values of the standard deviation mean a stable form factor. We grouped the pairs $(M^2,M'^2)$ by the same value of the standard deviation (contour lines inf Fig.\ref{fig1}). Once the different regions in the $M^2\times M'^2$ plane were identified, we tested the usual linear relation for the Borel masses, which leads to form factors with good stability \cite{bracco-2008-659,Rodrigues:2010ed}. This relation is given by: 
\begin{align}
M'^2 = \frac{m^2_o}{m^2_i}M^2\;.
\label{eq:ansatz}
\end{align}
We also considered a more  general linear relation between the Borel masses:
\begin{align}
M'^2 = a M^2 + b\;.
\label{eq:generalrelation}
\end{align}
In this last case, to test the stability of the form factor is necessary to determine 
the optimal $a$ and $b$ coefficients. In order to do this, we proceeded into an auto-consistent calculation: given a pair of values $(a,b)$ in Eq.(\ref{eq:generalrelation}), we tested many values of the Borel masses and thresholds satisfying conditions (i)-(ii), and calculated the associated standard deviation $\sigma^2(a,b)$. The optimal values of $(a,b)$ were found looking for the minimum of $\sigma^2(a,b)$ inside the studied $Q^2$ window.  

\section{Results and conclusions}
In the case of the $\rho$ off-shell diagram of the $D^* D \rho$ vertex, we found that the best linear relation between the Borel masses is $M'^2 = 0.1M^2 + 1.2$, with $\Delta_s = 0.3$ GeV and $\Delta_u = 0.6$ GeV. In Fig.\ref{fig1} we can see that this relation (dashed-dot line) crosses fewer contour lines than the usual relation $M'^2= m^2_D/m^2_{D^*}M^2$ of Eq.\ref{eq:ansatz}, showed also in Fig.\ref{fig1} (solid line). We obtain in this way a more stable form factor, as can be verified in Table \ref{tab1} looking at the standard deviation $\sigma^2$.
\begin{table}[h]
\begin{tabular}{c|cc|cc||cc|}
\cline{2-7}
 & \multicolumn{4}{c||}{$\Delta_s = 0.3\:\mbox{GeV}$, $\Delta_u = 0.6\:\mbox{GeV}$}  & \multicolumn{2}{c|}{$\Delta_s = \Delta_u = 0.5\:\mbox{GeV}$ \cite{Rodrigues:2010ed}} \\
 & \multicolumn{2}{c}{$M'^2 = 0.1M^2 + 1.2$} & \multicolumn{2}{c||}{$M'^2 = \frac{m^2_D}{m^2_{D^*}}M^2$}& \multicolumn{2}{c|}{$M'^2 = \frac{m^2_D}{m^2_{D^*}}M^2$}\\
\hline
  \multicolumn{1}{c}{$Q^2\:\mbox{(GeV$^2$)}$} & \multicolumn{1}{c}{$g^{(\rho)}_{D^*D\rho}(Q^2)$ } & \multicolumn{1}{c}{$\frac{\sigma_g}{(\bar{F}/100)} (\%)$ } & \multicolumn{1}{c}{$g^{(\rho)}_{D^*D\rho}(Q^2)$ } & \multicolumn{1}{c}{$\frac{\sigma_g}{(\bar{F}/100)} (\%)$ } & \multicolumn{1}{c}{$g^{(\rho)}_{D^*D\rho}(Q^2)$ } & \multicolumn{1}{c}{$\frac{\sigma_F}{(\bar{F}/100)} (\%)$} \\
\hline
        1.0  &  3.4277  & 13.56 & 3.5451 & 3.76 & 3.8176 & 4.13\\
        1.5  &  2.8002  &  5.84 & 3.0114 & 9.30 & 3.4245 & 12.34\\
        2.0  &  2.2737  &  4.73 & 2.5409 & 17.88 & 3.0295 & 19.66\\
        2.5  &  1.8343  &  9.31 & 2.1281 & 25.17 & 2.6511 & 25.85\\
\hline
\end{tabular}
\caption{Comparison between the $g^{\rho}_{D^*D\rho}(Q^2)$ and its standard deviation $\sigma$ obtained with the two relations for the Borel masses discussed in text. $\bar{F}$ is the mean value of the form factor.}
\label{tab1}
\end{table}

The coupling constant obtained with the new relation is $g^{(\rho)}_{D^* D \rho} = 6.68$ GeV$^{-1}$, which is $13\%$ bigger than the value of $g^{(\rho)}_{D^* D \rho} = 5.89$ GeV$^{-1}$  obtained with the relation of Eq.(\ref{eq:ansatz}) and also used in Ref.\cite{Rodrigues:2010ed}. This comparison was made with the same extrapolation used in  that reference. 
For the $B$ off-shell diagram of the $B_s^* B K$ vertex, we used $\Delta_s = 0.5$ GeV and $\Delta_u = 0.6$ GeV. In this case our study suggested the relation $M'^2 = 0.1M^2$, which leads to a more stable form factor, regarding the standard deviation, as can be verified in Fig.\ref{fig1} and Table \ref{tab2}.
\begin{table}[h]
\begin{tabular}{c|cc|cc||cc|}
\cline{2-7}
 & \multicolumn{4}{c||}{$\Delta_s = 0.5\:\mbox{GeV}$, $\Delta_u = 0.6\:\mbox{GeV}$} & \multicolumn{2}{c|}{$\Delta_s = 0.6\:\mbox{GeV}$,  $\Delta_u = 0.3\:\mbox{GeV}$ \cite{jr.:298}}\\
 & \multicolumn{2}{c}{$M'^2 = 0.1M^2$} & \multicolumn{2}{c||}{$M'^2 = \frac{m^2_\rho}{m^2_{B^*_s}}M^2$} & \multicolumn{2}{c|}{$M'^2 = \frac{m^2_\rho}{m^2_{B^*_s}}M^2$}\\
\hline
  \multicolumn{1}{c}{$Q^2 (\mbox{GeV}^2) $} & \multicolumn{1}{c}{$g^{(B)}_{B_s^* B K}(Q^2)$ } & \multicolumn{1}{c}{$\frac{\sigma_F}{(\bar{F}/100)} (\%)$} & \multicolumn{1}{c}{$g^{(B)}_{B_s^* B K}(Q^2)$ } & \multicolumn{1}{c}{$\frac{\sigma_F}{(\bar{F}/100)} (\%)$} & \multicolumn{1}{c}{$g^{(B)}_{B_s^* B K}(Q^2)$ } & \multicolumn{1}{c}{$\frac{\sigma_F}{(\bar{F}/100)} (\%)$} \\
\hline
        1.0  &  3.5569  &  2.04 & 2.5760  & 15.54 & 2.0083  &  3.66\\
        2.0  &  3.4556  &  2.01 & 2.5042  & 15.93 & 1.9498  &  3.31\\
        3.0  &  3.3598  &  2.04 & 2.4360  & 16.32 & 1.8943  &  3.02\\
        4.0  &  3.2691  &  2.11 & 2.3712  & 16.69 & 1.8414  &  2.80\\
\hline
\end{tabular}
\caption{Comparison between the $g^{(B)}_{B_s^* B K}(Q^2)$ and its standard deviation $\sigma$  obtained with the two relations for the Borel masses discussed in text. $\bar{F}$ is the mean value of the form factor.}
\label{tab2}
\end{table}

The coupling constant obtained with this new relation is $g^{(B)}_{B_s^* B K} = 20.0$, which is $27\%$ bigger than {the value of }$g^{(B)}_{B_s^* B K} = 15.7$, obtained with the relation $M'^2/M^2 = m^2_\rho/m^2_{B^*_s}$ of Ref. \cite{angelo,jr.:298} and with the same extrapolation of that reference.

Concluding, the results obtained in this work showed that a general linear relation between the Borel masses leads to more stable form factors. The results also suggest that the relation of Eqs.(\ref{eq:ansatz}) and (\ref{eq:generalrelation}) are, in fact, very similar.  This is true specially in the case of the $B_s^* B K$ vertex with the $B$ off-shell, for which panel (b) of Fig.\ref{fig1} shows that both relations give very similar results. For the $D^* D \rho$ vertex, the usual ansatz of Eq.(\ref{eq:ansatz}) gives results inside the stability region of the  $M'^2\times M^2$ plane, but is less stable when compare with Eq. (\ref{eq:generalrelation}), which crosses fewer contour lines. In addition, relation (\ref{eq:generalrelation}) does not respect the relation between the masses of the ingoing and outgoing mesons. Also the thresholds obtained when the stability of the form factor is imposed are different than the usual ones, of about $0.5$ GeV. These values, of the order of the gap between the ground and the first exited states, are required in order to obtain the correct mass and decaying coupling in two point QCDSR calculations.  
The study presented here was our first analysis about the question of the relations between Borel masses in QCDSR, and will be improved in a more thorough work.
\\
This work was sponsored by CAPES and CNPq.

\bibliography{PO-OsorioRodrigues}

\end{document}